\documentclass[
  journal=largetwo,
  manuscript=article-type,
  year=2020,
  volume=37,
]{cup-journal}

\usepackage{amsmath}
\usepackage[nopatch]{microtype}
\usepackage{booktabs}

\usepackage{tabularx}

\title{Verifying the Australian MWA EoR pipeline II: fundamental limits of the \pipe and the impact of instrumental effects}

\author{J.~L.~B.~Line}
\affiliation{International Centre for Radio Astronomy Research, Curtin University, Perth, WA 6102, Australia}
\alsoaffiliation{ARC Centre of Excellence for All Sky Astrophysics in 3 Dimensions (ASTRO-3D)}

\author{C.~M.~Trott}
\affiliation{International Centre for Radio Astronomy Research, Curtin University, Perth, WA 6102, Australia}
\alsoaffiliation{ARC Centre of Excellence for All Sky Astrophysics in 3 Dimensions (ASTRO-3D)}
\email[C.~M.~Trott]{cathryn.trott@curtin.edu.au}

\author{N.~Barry}
\affiliation{International Centre for Radio Astronomy Research, Curtin University, Perth, WA 6102, Australia}
\alsoaffiliation{ARC Centre of Excellence for All Sky Astrophysics in 3 Dimensions (ASTRO-3D)}

\author{D.~Null}
\affiliation{International Centre for Radio Astronomy Research, Curtin University, Perth, WA 6102, Australia}
\alsoaffiliation{ARC Centre of Excellence for All Sky Astrophysics in 3 Dimensions (ASTRO-3D)}

\author{C.~H.~Jordan}
\affiliation{International Centre for Radio Astronomy Research, Curtin University, Perth, WA 6102, Australia}
\alsoaffiliation{ARC Centre of Excellence for All Sky Astrophysics in 3 Dimensions (ASTRO-3D)}


\usepackage{xspace}

\newcommand{\hyperdrive}{\texttt{hyperdrive}\xspace}

\newcommand{\WODEN}{\texttt{WODEN}\xspace}
\newcommand{\pipe}{\texttt{AusEoRPipe}\xspace}

\newcommand{\twocm}{21-cm\xspace}

\usepackage{aas-macros}
\usepackage{hyperref} 
\hypersetup{colorlinks,citecolor=blue,linkcolor=blue,urlcolor=blue}

\begin{document}

\begin{abstract}
Detection of the weak cosmological signal from high-redshift hydrogen demands careful data analysis and an understanding of the full instrument signal chain. Here we use the \WODEN simulation pipeline to produce realistic data from the Murchison Widefield Array Epoch of Reionisation experiment, and test the effects of different instrumental systematics through the \pipe analysis pipeline. The simulations include a realistic full sky model, direction-independent calibration, and both random and systematic instrumental effects. Results are compared to matched real observations. We find that, (i) with a sky-based calibration and power spectrum approach we have need to subtract more than 90\% of all unresolved point source flux (10~mJy apparent) to recover \twocm signal in the absence of instrumental effects; (ii) when including diffuse emission in simulations, some $k$-modes cannot be accessed, leading to a need for some diffuse emission removal; (iii) the single greatest cause of leakage is an incomplete sky model; (iv) other sources of errors, such as cable reflections, flagged channels and gain errors, impart comparable systematic power  to one another, and less power than the incomplete skymodel.
\end{abstract}

\section{Introduction}
\label{sec:intro}
Exploration of the first billion years of the history of the Universe promises to shed light on the growth and evolution of the first generations of stars and galaxies, and the transformation of the cosmos from a neutral to a predominantly ionised IGM. One of the primary observational avenues for this period is the hyperfine transition from primordial neutral hydrogen gas that fills the IGM, which can be observed at low radio frequencies (50-200~MHz) with radio telescopes. The hydrogen brightness temperature encodes the radiation and thermal properties of the IGM over time and space; at early times, fluctuations are dominated by heating of the gas from the first stars and galaxies, while at later times, the lack of signal from ionised regions of hydrogen gas dominate the spatial fluctuations \citep{Furlanetto2006b}.

The cosmological \twocm signal is obscured by considerably brighter foreground emission from AGN and star forming galaxies, as well as our own Galaxy. Crucially, these are synchrotron and free-free emitters, with continuum spectra, allowing a spectral distinction to be made between them and the spectrally-structured \twocm line emission. This fundamental difference forms the basis for discriminating foreground contaminating power from the signal of interest for all experiments attempting this measurement \citep{koopmans2015,HERA2023,Beardsley2016,Trott2020,Barry2019,Patil2017,Mertens2020}. The brightest and closest of these foreground sources can be resolved and measured individually, and these form the basis for a sky-based approach to calibrating low frequency radio data. The calibration model and our understanding of the low-frequency radio sky are critical for obtaining clean and accurate data. In addition to careful calibration, there are other systematic effects that prevent a complete and pristine dataset. These include spectral channels flagged due to radio frequency interference \citep[][]{Wilensky2023}, time steps flagged due to RFI or instrument issues, ionosphere refraction of the signal, and an incomplete sky model, whereby some of the sky flux is missing from the calibration model \citep{Barry2019}. These act to make both calibration data and science data inaccurate. Understanding the impact of these effects is crucial for (i) obtaining the cleanest set of data, (ii) prioritising effort to address particular systematics, (iii) confidence in the robustness of the reported \twocm signal power. In \citet{Line2024}, the end-to-end data processing pipeline was tested to ensure there is no signal loss from our techniques. In this paper, realistic simulations are used to test the impact of different random and systematic errors and provide guidance on their importance for the next generation of analysis tools.

In Paper I of this series~\citep{Line2024}, the \pipe was tested against signal loss with a model \twocm signal, ensuring that the methodology does not bias the signal power. In this paper, we focus attention on the impact of calibration, and instrumental systematics on the ability to detect the \twocm signal. We focus on the MWA EoR high-band frequency range (167 - 198MHz), as this is covered by the \twocm sky model (detailed in Section~\ref{subsec:21cm_model}). We focus on the MWA phase I layout. Due to computational constraints, we limit ourselves to the EoR0 field (centred at RA,Dec = $0^h, -30^{\circ}$). We only consider Stokes I sky models as this not only reduces computational costs, but the calibration catalogue used by the \pipe is already Stokes I only. We further constrain ourselves to zenith beam-pointings; in doing so, we have the computational resources to simulate multiple observations closely spaced in LST. This allows us to experiment with averaging calibration solutions, as the MWA instrument should be somewhat stable over the space of half an hour (see Section~\ref{sec:calib_averaging} for averaging results). Even with these constraints, we are able to perform simulations that test the fundamental limits of \twocm recovery with the \pipe in the presence of foregrounds, as well as the impact of instrumental effects on the \pipe (Section~\ref{sec:inst_effects}).

The paper is structured as follows. In Section~\ref{sec:overview}, we overview the \pipe and strategy for simulating visibilities. In Section~\ref{sec:skymodel} we detail the point, diffuse, and \twocm sky models used for simulating visibilities. In Section~\ref{sec:pristine_sim} we investigate the effects of calibration and subtraction on simulated data containing no instrumental effects; we add instrumental effects in Section~\ref{sec:inst_effects}. In Section~\ref{sec:calib_averaging} we explore averaging calibration solutions over time. Finally, we discuss and conclude our work in Sections~\ref{sec:discuss} and ~\ref{sec:conclusion}.

\section{Pipeline and Simulation strategy}
\label{sec:overview}

\subsection{\pipe overview}
\label{subsec:pipeline overview}
For a full overview of the \pipe, including various software packages, see~\citep{Line2024}. Here we include the most pertinent details.

MWA observations are undertaken in 2-min snapshots, sampled at 2$\,$s and 40$\,$kHz resolution. Direction-independent calibration is applied at native resolution, before the data are averaged to 8$\,$s and 80$\,$kHz for further analysis. During the 2-min snapshot, the phase centre is fixed in celestial coordinates, and the beam pointing centre is fixed, meaning the sky drifts a small amount through the primary beam. The calibration is performed on a per-channel basis. This choice for the vanilla version of the \hyperdrive software was made in response to the spectral polynomial fitting that was performed by the earlier RTS software \citep{Mitchell2008}, which imparted systematic power into the power spectrum (PS), and could not handle cable reflections. In this work, we maintain the per-channel calibration strategy and present the results within that framework.

Direction-independent calibration uses a sky model constructed from unresolved (point) and extended sources that are above the horizon at the observation time. The sky model is constructed from a combination of GLEAM \citep{Wayth2015,Hurley-Walker2017} and LoBES \citep{Lynch2021} for point and multi-component sources, and additional custom models for A-Team radio galaxies and supernova remnants \citep{Cook2022,Line2020}. Within the EoR fields, this catalogue is 90\% complete to 32 mJy \citep{Lynch2021}. Notably, the calibration sky model does not include diffuse emission. This omission is handled by only calibrating with baselines longer than 30$\lambda$, where the diffuse power is sub-dominant. Nonetheless, this is a shortcoming of the calibration sky model. The same sky model can then be used for foreground subtraction, whereby model visibilities formed from the catalogue are directly subtracted from the calibrated visibilities. This is a good path for removing foreground power, but assumes that there are no direction-dependent (DD) effects such that the model deviates from the measurements. DD calibration, such as peeling \citep{Mitchell2008}, is not implemented for this work.


\subsection{Simulation strategy}

We make one further concession to computational and development time, and only focus on instrumental effects that are independent of direction upon the sky. These kinds of effects, such as visibility flagging and internal cable reflections, can be added to visibilities post simulation. In this way, a base set of simulations can be used to test a number of instrumental effects in isolation, for little extra compute. Sky directional effects, such as the ionosphere and directional RFI, must be simulated during the calculations of the visibilities themselves. This is outside the scope of this paper, and left for future work.

\subsection{Data selection}

In all \pipe testing, we aim to reproduce as many real effects as possible. As such, we simulate with observational settings matching real MWA data for comparison. We select a set of 15 EoR0 zenith pointings from 2015, spanning observation GPS times 1125938488 to 1125940192 (UTC 2015-09-10 16:41:11 - UTC 2015-09-10 17:09:35). Note that GPS start time is used for the observation identifier within the MWA. These observations are known to produce good calibration solutions and have little ionospheric activity \citep{Jordan2017}, allowing for a reasonable comparison to simulation. We only simulate a single pointing (zenith) to reduce computational load, particularly with the diffuse and \twocm emission.{  In \citet{Trott2020}, the behaviour of individual versus combined pointings in the PS did not show significant differences, justifying use of a single pointing for this work. We simulate all data at a 2s, 40kHz to match the resolution of real data, ensuring any effects of averaging to 8s, 80kHz when applying calibration solutions or creating power spectra are captured. All power spectra shown come from the north-south aligned dipoles, because these are shown to consistently produce cleaner power spectra in previous MWA publications with this pipeline \citep{Trott2016}. A weighted FFT is used throughout for spectral analysis. }

\section{Sky models}
\label{sec:skymodel}
Broadly there are three regimes of sky signal we are interested in. The first is the \twocm signal itself. The second we describe as discrete (compact) sources. We define any singular astrophysical object as a discrete source, and so include extended A-team sources here. These are sources that contribute sky power on smaller angular scales, and can be used as calibrator sources. The third we call the diffuse, which broadly covers all large angular-scale emission, predominantly coming from synchrotron emission from the Milky Way. Both the discrete and diffuse affect calibration and signal recovery in different ways, so including both is paramount to pipeline testing. We can take advantage of the additive nature of visibilities, allowing us to simulate a base set of \twocm visibilities, and add on the two foreground models in stages to test these differences in isolation.

The three sky models used in the work are detailed in the following subsections. All are visualised as seen from the MWA when the EoR0 field is transiting the meridian in Figure~\ref{fig:all_skymodels}.

\subsection{\twocm model}
\label{subsec:21cm_model}
We use the \twocm model described in Paper I \citep{Line2024}, which was designed to match EoR0 high-band. It covers a $\sim 50\times50$ squared degree field, at a spectral resolution of 80kHz, and angular resolution of $\sim27\,$arcsecs. The model is a TAN FITS projection with each pixel represented as a point source in the sky model, with the $\sim27\,$arcsec resolution over-sampling the $\sim2\,$arcmin MWA resolution. The visibilities are analytically predicted from a point source via assuming a point source is a dirac-delta function and applying that to the measurement equation. For the exact mechanics of the \WODEN simulator, please see~\citep{Line2022}. For details of how the \twocm model was projected and interpolated onto the sky, along with the ability of the \pipe to recover the expected PS, please see \citep{Line2024}.

\subsection{Discrete foreground model}
\label{subsec:discrete_model}
\begin{figure*}[h!]
  \centering
  \includegraphics[width=\columnwidth]{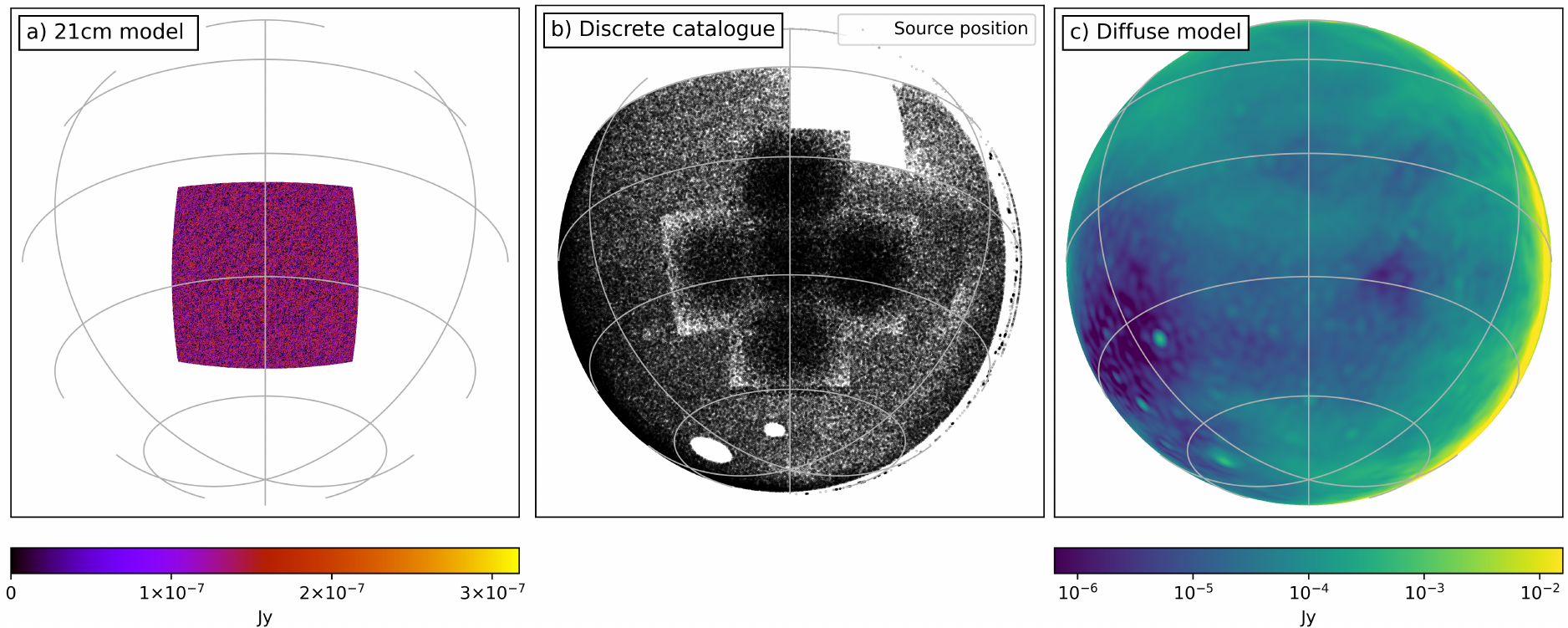}
  \caption{All-sky orthographic projections of the sky models, centred at RA,Dec = $0^{\circ},-27^{\circ}$ (-27$^{\circ}$ is zenith for the MWA), where: $a)$ shows a slice of the \twocm sky model at 167MHz; $b)$ shows the positions of all sources in the discrete sky model; $c)$ shows the diffuse sky model at 200MHz}
  \label{fig:all_skymodels}
\end{figure*}

For this model we use the current calibration catalogue as used by the \pipe. This hybrid catalogue uses GLEAM as a base~\citep{Hurley-Walker2017}, with LoBES~\citep{Lynch2021} covering the EoR0 field, and the work of \citet{Procopio2017} covering the EoR1 field. LoBES is used to supplant the EoR0 field as it was created with the MWA Phase II array layout, offering a resolution of $\sim80\,$arcsec, compared to the $\sim2\,$arcmin resolution of GLEAM, which was created with the MWA phase I layout. As a result, LoBES is 90\% complete at 32$\,$mJy whereas GLEAM is 90\% complete at 170$\,$mJy. This difference is visualised in the denser central fields in Figure~\ref{fig:all_skymodels}{\color{blue}b}.

A number of extended sources are modelled including Fornax A~\citep{Line2022}, and Galactic supernova remnants~\citep{Cook2022}. All source spectral energy distributions are modelled by either a power-law or curved power-law model, making all discrete sources spectrally smooth. {Most sources are represented as point sources in the sky model, with a number Gaussian components used for sources with more structure on the sky. The most complicated sources are represented by Shapelets, including Fornax A. Again, all visibilities are analytically calculated from the point source, Gaussian, or Shapelet parameters. The visibility calculations for a Gaussian and Shapelet component are again detailed in~\citep{Line2022}. }Figure~\ref{fig:all_skymodels}{\color{blue}b} shows the sky coverage of this model; the missing area in the top right is due to the underlying GLEAM catalogue coverage. { In total, the discrete sky model has 338,797 sources.}

\subsection{Diffuse foreground model}
\label{subsec:diffuse_model}
The diffuse sky model is based on an $m$-mode analysis map as described in~\citet{Kriele2022}, made using the Engineering Development Array 2~\citep[EDA2][]{Wayth2022}. The EDA2 is a low angular resolution telescope based at the same site as the MWA, and using the same MWA dipoles, making the sky coverage perfectly matched for our purposes. The original Stokes I map was imaged at 159~MHz into a HEALPix~\citep{Gorski2005} $N_{\mathrm{side}}=64$ projection (a HEALPixel resolution of $\sim0.9^\circ$). \citet{Kriele2022} also produced an accompanying spectral index map. To match the resolution of the MWA, we take the original Stokes I and spectral index maps, and upgrade to $N_{\mathrm{side}}=2048$ ($\sim1.7\,$arcmins) using the \texttt{healpy}~\citep{Zonca2019} \texttt{ud\_{}grade} function. We then smooth the map using the \texttt{healpy.smoothing} function with a $0.9^\circ$ kernel, to remove the edges of the original HEALPixels. It should be noted that this diffuse map includes all sky emission, meaning all discrete sources as described in ~\ref{subsec:discrete_model} are present. The consequence being any simulation containing both sky models will result in some double counting of flux. However, for the purposes of this paper, both models combined need not perfectly match the sky; as long as they produce comparable power to that seen in real data, they can be used in pipeline validation.

\section{Pristine simulation}
\label{sec:pristine_sim}
\begin{figure*}[h!]
  \centering
  \includegraphics[width=0.8\columnwidth]{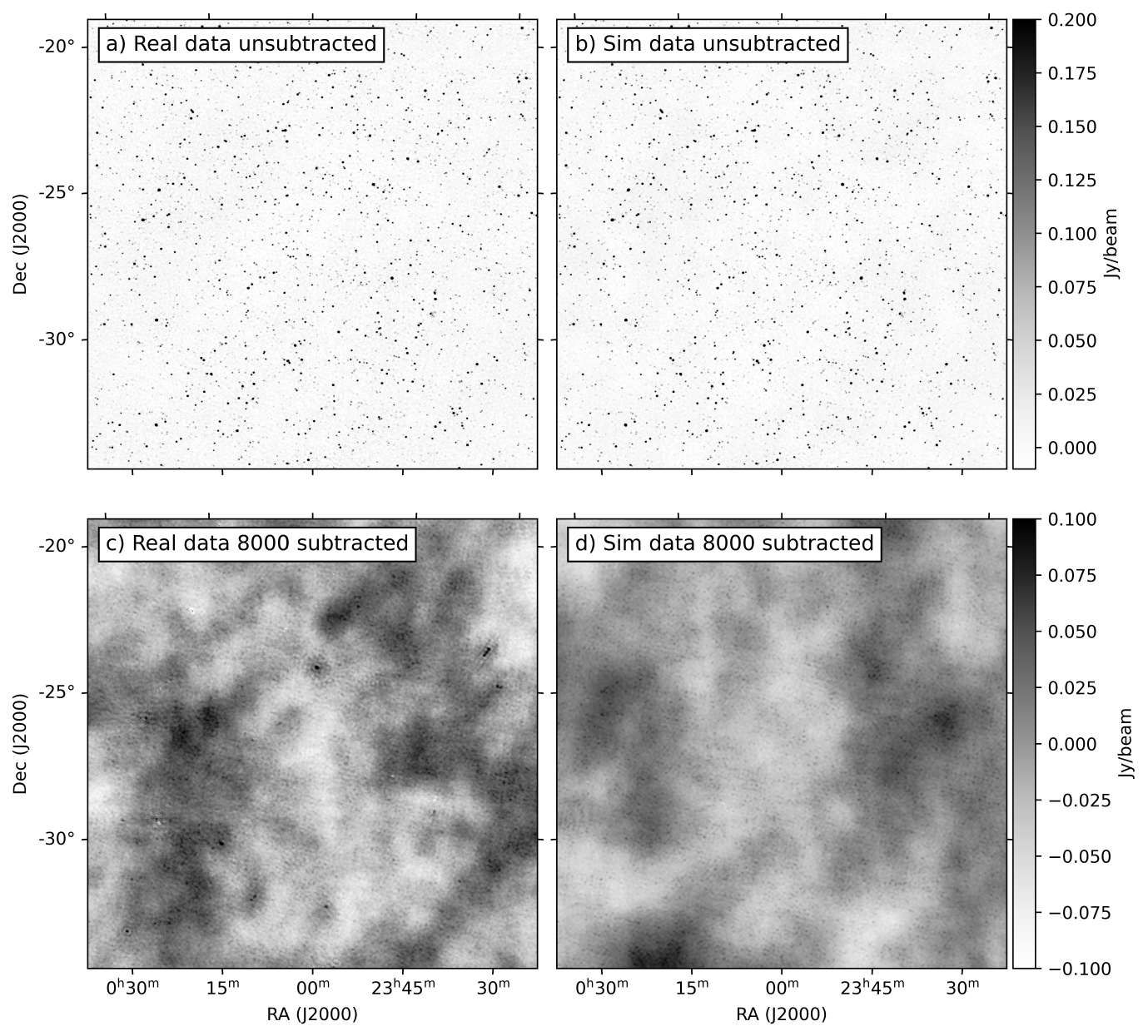}
  \caption{Comparison of a real zenith pointed 2 minute snapshot to simulated data. Both real and simulated data were averaged to 8~s, 80~kHz, and the full bandwidth imaged via \texttt{WSClean}. The top row were both imaged with Briggs 0 weighting, where \textit{a)} shows the real data after calibration, and \textit{b)} shows the simulated data with no calibration. The bottom row were both imaged with natural weighting, where \textit{c)} shows the real data after subtracting 8000 sources, and \textit{d)} shows the simulated data with the same 8000 sources subtracted.}
  \label{fig:compare_sim_to_real_sky}
\end{figure*}

We first test the \pipe in the presence of foregrounds, without systematic instrumental errors. We simulate a single EoR0 zenith observation 1125939344, which has an LST$=359.9^{\circ}$, setting the EoR0 field close to directly at zenith. We simulate all three sky models, and check the ability of the \pipe in recovering the \twocm with and without calibration involved. After combining all three sky models, we image the visibilities using \texttt{WSClean}~\citep{Offringa2017, Offringa2019} and compare to real data in Figure~\ref{fig:compare_sim_to_real_sky}. To compare the diffuse model, we subtract the same 8000 discrete sources from the real and simulated data. This should leave the similar amounts of discrete power in both the real and simulated data, down to the completion level of the LoBES catalogue. This qualitative comparison shows excellent agreement with the discrete model to real data. The diffuse model does not align perfectly, but shows similar power on the sky, as well as angular distribution of that power.

\subsection{Direct subtraction}
\label{subsec:pristine_direcsub}

In this subsection we test how much discrete sky model flux must be subtracted before we can recover the \twocm signal. { We start by creating a two minute simulation containing the full discrete and \twocm sky models as a visibility test bed. With the EoR0 field centre at zenith, there are 227,585 discrete sources above the horizon, all of which are present in the test bed visibilities. We then generate three sub skymodels to subtract from the test bed, by cutting the sky model at different flux thresholds ($10^{-1},\,10^{-2},\,10^{-3}\,$Jy). These flux thresholds are in reference to the apparent flux, where we have weighted the discrete sky model by the primary beam to calculate the apparent flux of all sources at the central frequency of 182~MHz. See Table~\ref{table:flux_cuts} for the resultant numbers after these cuts were applied.}

\begin{figure}[h!]
  \centering
  \includegraphics[width=\columnwidth]{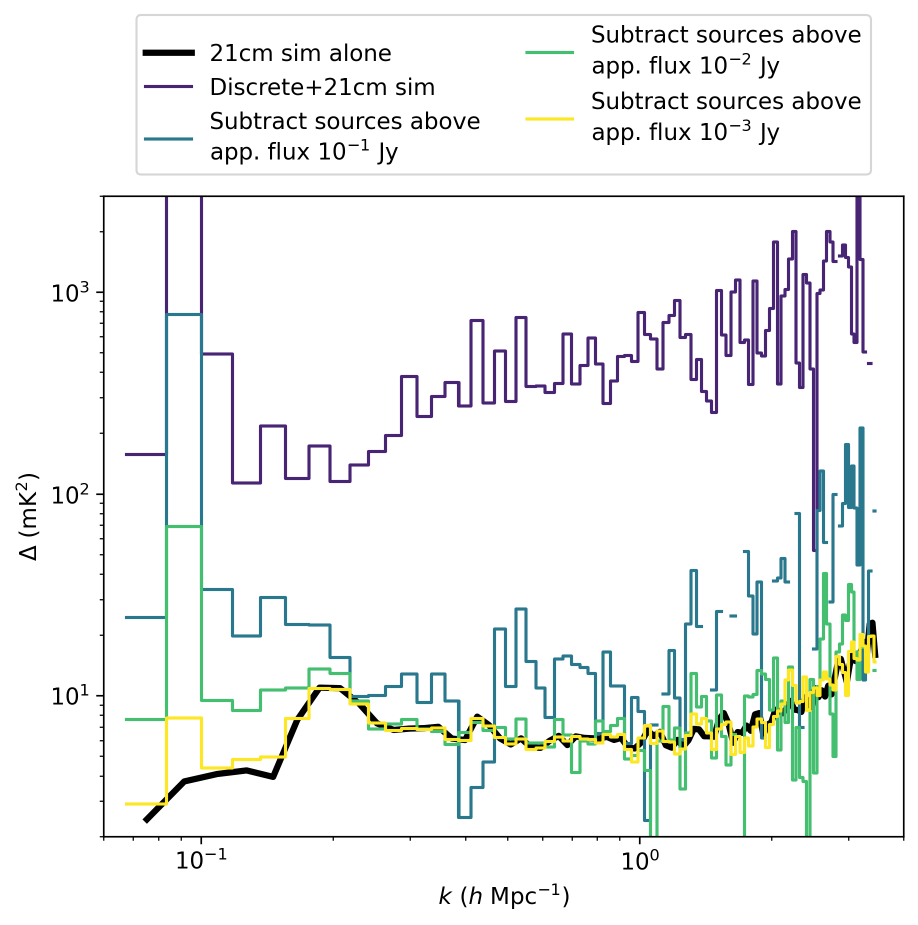}
  \caption{1D PS depicting data from a single zenith EoR0 observation where only the discrete foregrounds were included are shown. All PS were obtained after performing a wedge cut.}
  \label{fig:nocal_discrete_subtraction}
\end{figure}

We simulate these three sub sky models through \WODEN and then subtract them from the test bed directly in visibility space. Resultant 1D PS are shown in Figure~\ref{fig:nocal_discrete_subtraction}. { Before averaging from a 2D to a 1D PS, the foreground wedge is avoided by excluding any modes where $k_\parallel < 0.08 \, h \textrm{Mpc}^{-1}$, $k_\perp > 0.06 \,h \textrm{Mpc}^{-1}$, and performing a horizon line cut (discarding any modes in the wedge as marked by the solid black line as shown in Figure~\ref{fig:comp_avg_to_normal_cal_2D}.  }Figure~\ref{fig:nocal_discrete_subtraction} shows that the apparent flux cut at $10^{-3}$Jy is necessary to reliably recover the \twocm signal.

Note this is a flux cut on this particular discrete sky model; this is not saying that leaving all sources below $10^{-3}$Jy in real data will allow a detection. This result is limited by the completeness of this discrete sky model, and these simulations do not include confusion noise.{  It should also be noted that as we cut by apparent flux, at the lower the flux cut thresholds, we are adding more sources outside the main lobe of the MWA primary beam. These sources that are further from field centre contribute to higher $k_\parallel$ values due to projection effects. So although they contribute less overall power than sources in the main primary beam lobe, they contribute power closer to the so-called 2D PS "window", an area of the PS expected to have the least contamination from foreground sources.}

\begin{table}[h]
\centering
\renewcommand{\arraystretch}{1.0}
\caption{Cuts placed on discrete sky model and resultant number of sources { and their contribution to the apparent flux. We include a column detailing the total number of sources above the horizon without a flux cut for reference.}}
\begin{tabularx}{.99\columnwidth}{>{\setlength\hsize{.2\hsize}} X | >{\setlength\hsize{.2\hsize}} X | >{\setlength\hsize{.3\hsize}} X | >{\setlength\hsize{.3\hsize}} X}
\hline
{Apparent flux cut (Jy)} & {Num. sources} & {Fraction of total apparent flux (\%)} & {Fraction of total num. sources (\%)} \\
\hline
\hline
$10^{-1}$ & 3027 & 55.3 & 1.3 \\
$10^{-2}$ & 26359 & 88.6 & 11.6 \\
$10^{-3}$ & 79766 & 98.5 & 35.0 \\
No cut & 227585 & 100.0 & 100.0 \\
\hline
\end{tabularx}
\label{table:flux_cuts}
\end{table}

\begin{figure}[h!]
  \centering
  \includegraphics[width=\columnwidth]{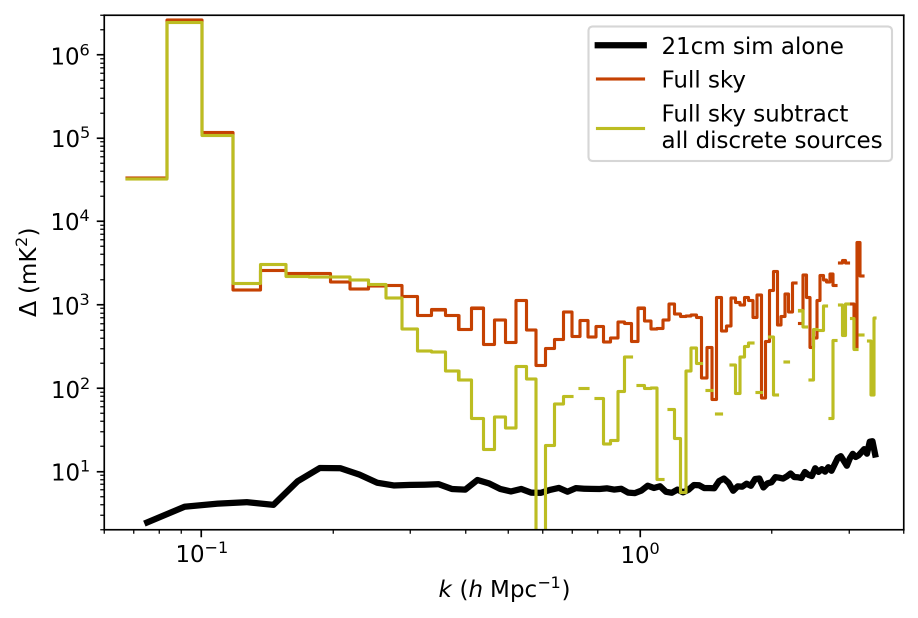}
  \caption{1D PS depicting data from a single zenith EoR0 observation where both discrete and diffuse foregrounds were included.}
  \label{fig:nocal_diffuse_subtraction}
\end{figure}

We repeated this experiment after adding the diffuse sky model to the test bed. We call this combination of \twocm, discrete, and diffuse sky models as the full sky model. We find that subtracting the three sub discrete sky models makes little difference to the results, and instead subtract the entire discrete sky model. Results are shown in Figure~\ref{fig:nocal_diffuse_subtraction}. As mentioned in Section~\ref{subsec:diffuse_model}, the diffuse sky model contains the entire visible radio sky, and so may contain duplicate flux at higher $k$-modes (small angular scales). Residual power left at high $k$-modes after subtraction therefore may be false power. However, the large residual power seen at lower $k$-modes comes exclusively from the diffuse model. These results show for this single observation, the \twocm signal is unrecoverable at low $k$-modes without some removal of power from the diffuse foregrounds.

\subsection{Calibration and subtraction}
\label{subsec:pristine_calsub}
\begin{figure}[h!]
  \centering
  \includegraphics[width=\columnwidth]{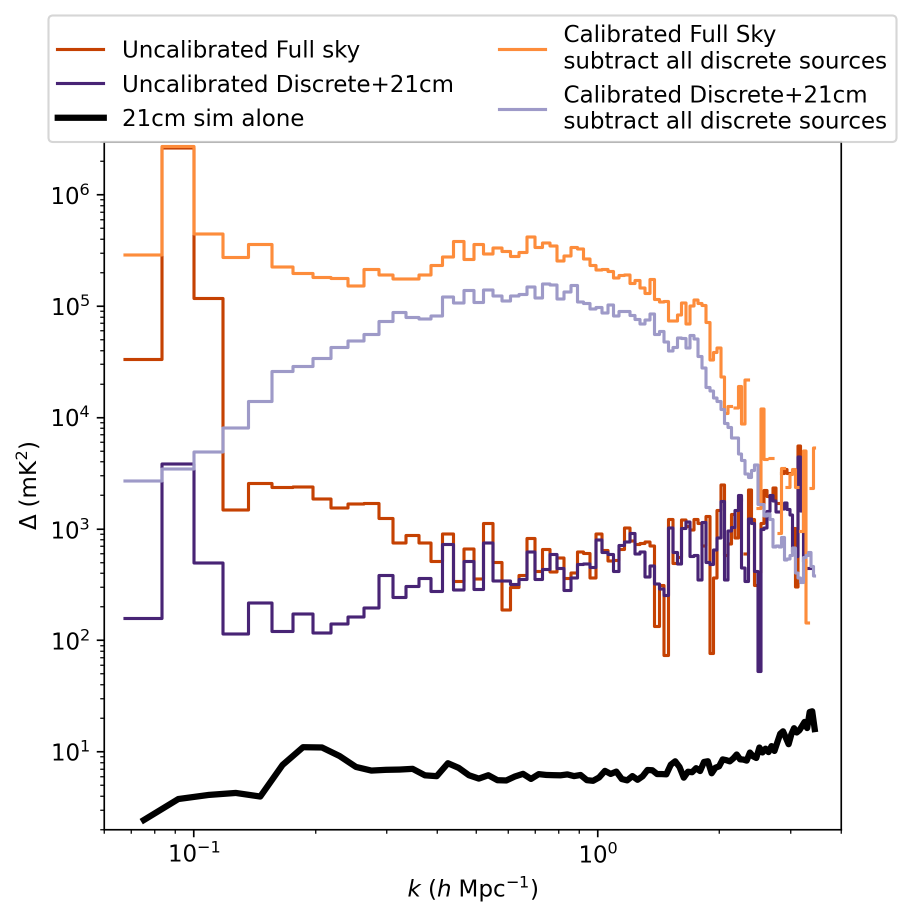}
  \caption{1D PS depicting data from a single zenith EoR0 observation, showing the effects of calibration on leakage into the window. Note that the dark purple and brown lines showing uncalibrated simulations are colour-coded to lines showing the same simulations in Figures~\ref{fig:nocal_discrete_subtraction} and~\ref{fig:nocal_diffuse_subtraction}, for easy comparison.}
  \label{fig:cal_diffuse_subtraction}
\end{figure}

We repeat the subtraction experiment in Section~\ref{subsec:pristine_direcsub} but now include calibration. We use \hyperdrive \texttt{di-calibrate} to calibrate the full sky model visibilities, using a calibration catalogue of the apparent brightest 8000 sources in the discrete sky model. We then use \hyperdrive \texttt{solutions-apply} to apply those calibration solutions to the full sky model visibilities. We then subtract the uncalibrated discrete sky model visibilities from the calibrated visibilities\footnote{This last step is equivalent to using \hyperdrive \texttt{vis-subtract} using the entire calibration catalogue. We re-use existing \WODEN simulations to save on compute.}. To summarise, we calibrate using an incomplete discrete sky model, and then subtract the entire discrete sky model from the visibilities, { thereby including a systematic error in the calibration model.} We also repeat this procedure using a combination of just the discrete and \twocm sky models. If we had perfect calibration, we would expect this to result in a perfect recovery of the \twocm signal. Results are shown in Figure~\ref{fig:cal_diffuse_subtraction}. These results show that calibration induces up to three orders of magnitude of power into the window, when using a single observation. This comes from percent-level amplitude fluctuations in the calibration solutions, that vary quickly as a function of frequency. This couples power at low $k$-modes from the spectrally-smooth foregrounds, into higher $k$-modes. This leakage completely masks the \twocm signal. We note here that applying the calibration solutions to the \twocm simulation alone does not bias the recovery of the signal. Only a small fraction of power in the sky is perturbed, and so the \twocm alone is greater than the leakage caused by calibration solutions from itself.

\section{Instrumental Effects}
\label{sec:inst_effects}
In this Section we detail the direction independent{  random and systematic errors and how they are simulated. We use a test bed of 30 minutes of EoR zenith simulations (15 contiguous 2 minute snapshots)}, containing both the diffuse and discrete sky models. We apply each error in isolation to this test bed, and investigate the ability of \hyperdrive to calibrate them in subsection~\ref{subsec:apply_inst_errors}. All instrumental effects were added using \texttt{add\_{}instrumental\_{}effects\_{}woden.py}\footnote{See~\url{https://woden.readthedocs.io/en/latest/scripts/add_instrumental_effects_woden.html} for documentation.}

\subsection{Edge and centre channel flagging}
The legacy MWA correlator used a polyphase filterbank \citep{Tingay2013} to frequency channelise data. The entire bandwidth was split across 24 "coarse bands", each of 1.28MHz bandwidth, with a frequency-dependent bandpass imparted by the filterbank. These coarse bands suffered from low SNR at the edges, and aliasing causes the central channel to also degrade. For this reason, calibration is more effective when flagging these channels. Given we are testing 40kHz resolution data, the first systematic instrumental effect is to simply flag the first two, central, and final two channels in each coarse band\footnote{This is easily achieved with \hyperdrive by passing the optional argument \texttt{--fine-chan-flags-per-coarse-chan 0 1 16 30 31}}. These regular spectral flags cause harmonic modes, linking foreground power from low to high $k_\parallel$. The current version of the CHIPS PS processing software can optionally include a non-uniform FFT calculator to handle these missing channels. Besides flagging for bandpass effects, real data are also flagged for malfunctioning receiving elements and RFI. We leave investigating these flagging effects for future work.

\subsection{Tile based gain error}
Each receiving element (often called a "tile" for MWA,) in a dual polarisation interferometer can have a gain ($g_x, g_y$) and leakage ($D_x,D_y$) term for each polarisation. Each visibility is a correlation of two tiles. \WODEN implements any tile gains and leakages by multiplying the visibility between tiles 1 and 2 through the following operation:
\begin{multline}
  \begin{bmatrix}
  V^{\prime}_{12\,XX} & V^{\prime}_{12\,XY} \\
  V^{\prime}_{12\,YX} & V^{\prime}_{12\,YY}
  \end{bmatrix} = \\
  \begin{bmatrix}
  g_{x1} & g_{x1}D_{x1} \\
  g_{y1}D_{y1} & g_{y1}
  \end{bmatrix}
  \begin{bmatrix}
  V_{12\,XX} & V_{12\,XY} \\
  V_{12\,YX} & V_{12\,YY}
  \end{bmatrix}
  \begin{bmatrix}
  g_{x2}^{\ast} & g_{x2}D_{x2}^{\ast} \\
  g_{y2}D_{y2}^{\ast} & g_{y2}^{\ast}
  \end{bmatrix}^T
\end{multline}
where $V$ is a visibility, $\ast$ means the complex conjugate, $T$ is the transpose, and $V^{\prime}$ is the visibility after the tile gains and leakages have been applied. The leakage terms are calculated via Equation A4.5 from~\citet{TMS3}:
\begin{align}
D_x = \Psi - i \chi \\
D_y = -\Psi + i \chi,
\end{align}
where $\Psi, \chi$ are alignment errors of the dipoles. This equation is really designed for single antennas, but in the MWA case, one could assume all dipoles in a tile are aligned perfectly to the mesh, and the mesh is slightly offset. This would mean the alignment errors for all dipoles are the same.

In this work, we apply a different random gain error to each tile and polarisation. We draw the amplitudes of $g_x, g_y$ from a uniform distribution between 0.7 and 1.3, $U(0.7,1.3)$. We add a phase slope to $g_x, g_y$ as a function of frequency with a maximum phase offset drawn from $U(-60^{\circ},+60^{\circ})$. We draw $\Psi$ from $U(0,0.02^{\circ})$ and $\chi$ from $U(0,0.05^{\circ})$. 
The random gain errors are kept constant across all 15 observations, { and are therefore a systematic error}. The manifestation of the gain errors can be seen in Figure~\ref{appfig:gain_calsols}.

\subsection{Cable reflections}
\label{subsec:cable_reflect}
Mismatched impedance at coaxial cable ends can cause internal reflections that setup standing waves, adding frequency-dependent ripples to the visibilities. We follow the formalism from~\citet{Beardsley2016} to define the cable reflection gain seen by tile $i$ for polarisation $\mathrm{pol}$ as
\begin{equation}
  R_{\mathrm{pol},i}(\nu) = R_{0,i} \exp(-2\pi i \nu \tau_i),  
\end{equation}
where $R_{0,i}$ is the complex reflection coefficient, and $\tau_i$ is the time delay caused by the cable length connected to tile $i$. The time delay is given by
\begin{equation}
\tau_i = \frac{2l_i}{0.81c},
\end{equation}
where $l_i$ is the length of the cable connected to tile $i$, and $c$ is the speed of light. The factor 0.81 comes from the velocity factor of the cable, which we again take from~\citet{Beardsley2016}.

The amplitudes of these cable reflections have been measured to average between 0.02 and 0.1 using \texttt{FHD} (Barry, private communication). We therefore draw the amplitude of $R_{0,i}$ from $U(0.02, 0.1)$, and add a random phase offset of $U(0.0, 180^{\circ})$ for each tile.

\subsection{Noise}
Thermal noise on cross-correlations from an interferometer are estimated from both internal receiver temperature ($T_{\mathrm{rec}}$) and the sky temperature ($T_{\mathrm{sky}}$), via Equation 6.50 in~\citet{TMS3}:
\begin{equation}
\sigma_{\mathrm{cross}} = \frac{\sqrt{2}k_b(T_{\mathrm{sky}} + T_{\mathrm{rec}})}{A_{eff}\sqrt{\Delta\nu\Delta t}},
\end{equation}
where $A_{eff}$ is the effective area of the tile, $\Delta\nu$ is the channel width, and $\Delta t$ is the integration time. $\sigma_{\mathrm{cross}}$ describes the standard deviation of a zero mean Gaussian noise distribution. Throughout this paper, we set $T_{\mathrm{sky}} = 228\,\textrm{K}, T_{\mathrm{rec}} = 50\,\textrm{K}, A_{eff} = 20.35\,\textrm{m}^2$. A different realisation of the noise is added to each simulated observation, { inducing a random error.}


\subsection{Applying instrumental effects}
\label{subsec:apply_inst_errors}
\begin{figure*}[h!]
  \centering
  \includegraphics[width=0.9\columnwidth]{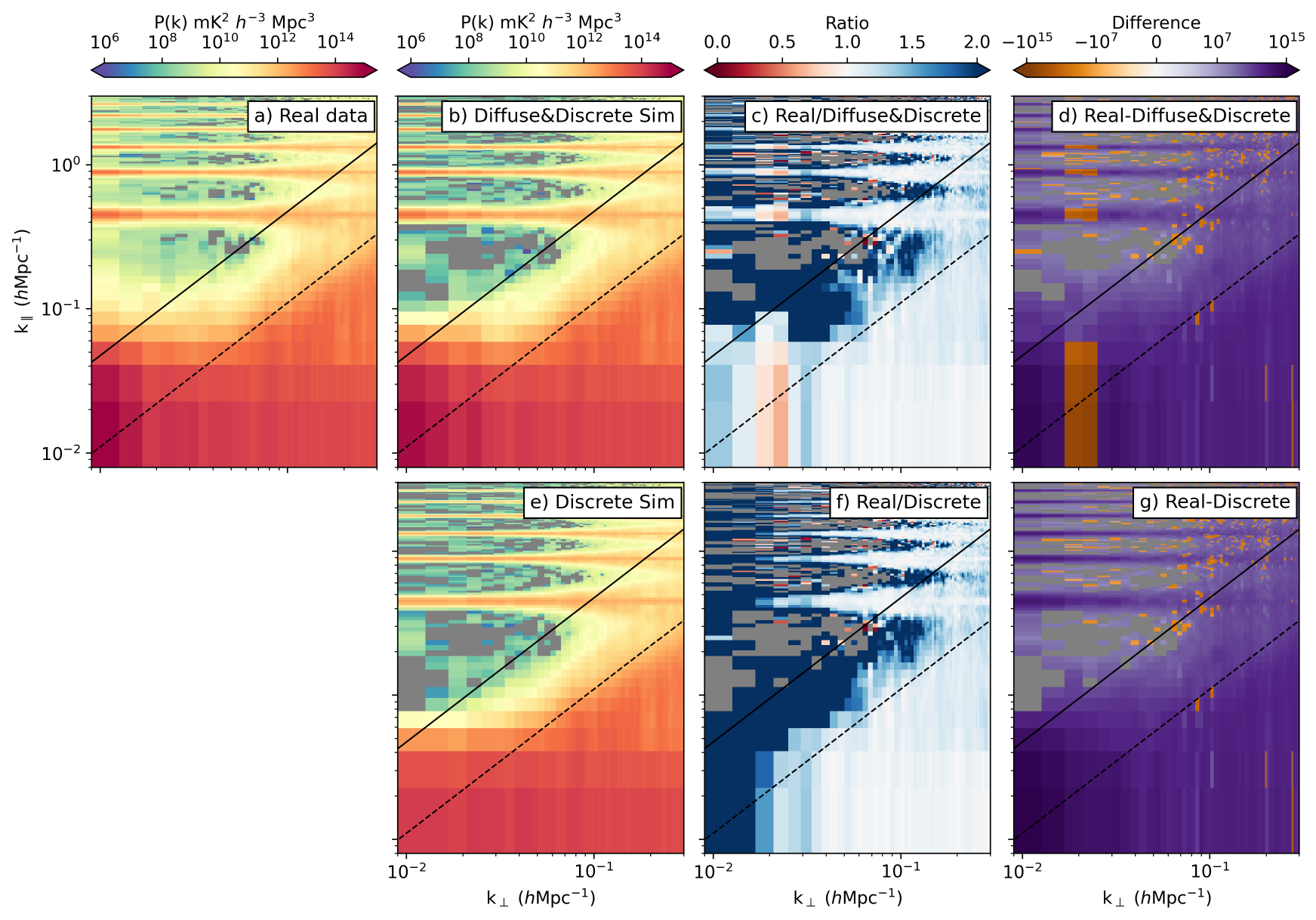}
  \caption{Comparison of an integration over 15 real two-minute zenith observations to simulated data. All real and simulated data have been calibrated using 10,000 sources, with the real data calibrated using 8000 sources. In the ratios, blue means more power in the real data, red means less power. In the differences, purple means more power in the real data, and orange less. Negative pixels in \textit{a), b), e)} are shown in grey. These pixels have also been masked in the ratio and differences.}
  \label{fig:comp_real-to-sim_2D}
\end{figure*}

We take each effect detailed in Section~\ref{sec:inst_effects} and apply them to the 15 zenith observations in isolation, and all in combination. We calibrate all simulations using the apparent 10,000 brightest sources in the sky for the given LST of that simulation. We integrate all 15 observations containing all instrumental effects into 2D PS and compare to real data in Figure~\ref{fig:comp_real-to-sim_2D}. We compare simulations containing both the discrete and diffuse sky models, as well as only the discrete sky model. Qualitatively it can be seen from the top row that the added instrumental errors match the real data, given the matching power in coarse band harmonics, and leakage at $k_\parallel \sim 0.8$. However, there is clearly less leakage into the window from the simulations, indicating further unmodelled systematics. As expected, Figure~\ref{fig:comp_real-to-sim_2D} shows missing power at the large spatial scales and along the horizon when only simulating the discrete sky model. \hyperdrive calibrates cable reflections well due to its per-channel calibration strategy.

\begin{figure*}[h!]
  \centering
  \includegraphics[width=\columnwidth]{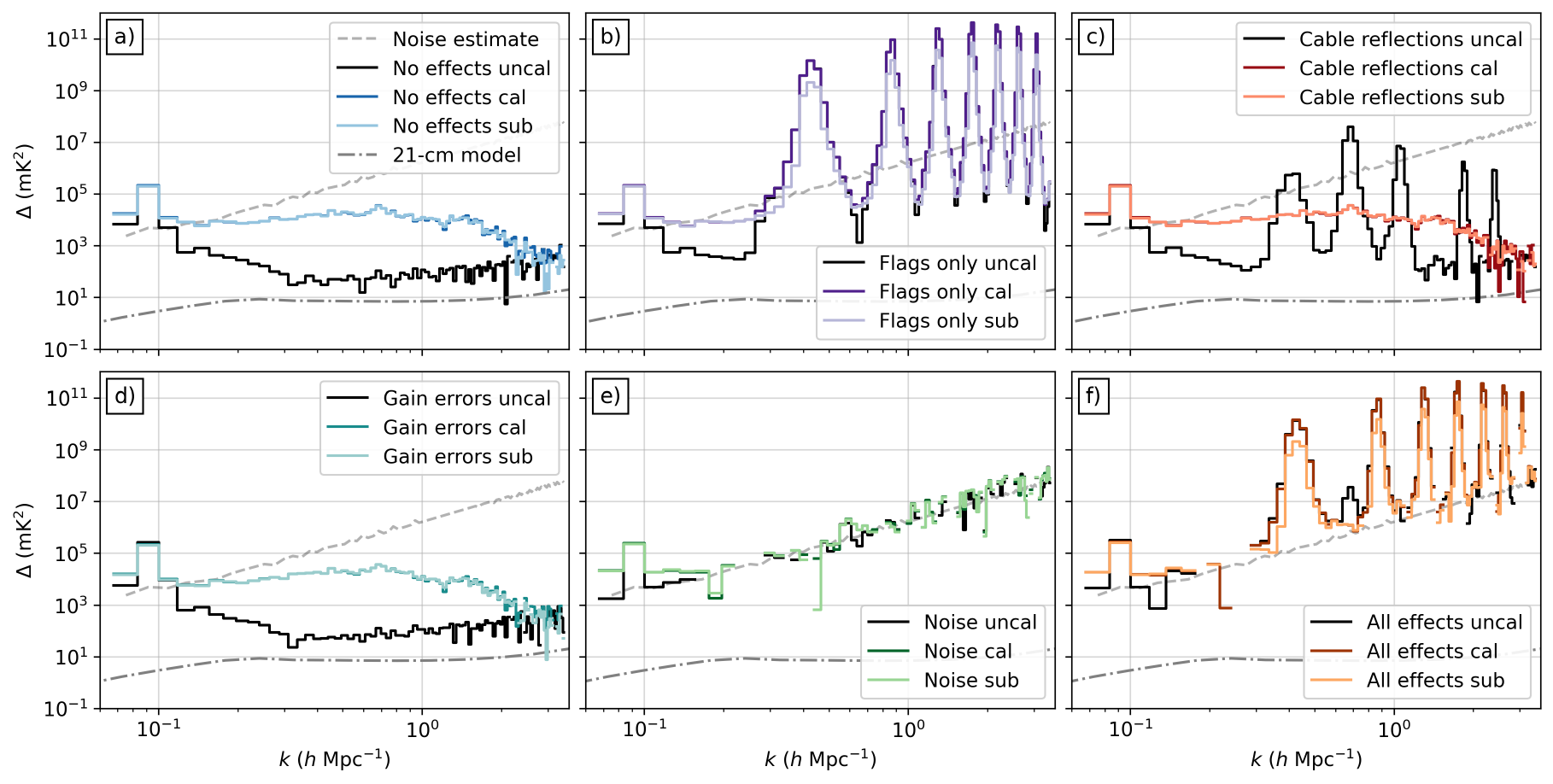}
  \caption{Comparison of different instrumental effects and their manifestation in the window. These 1D PS were made from 15 zenith observations, the 2D PS of which are shown in Figure~\ref{fig:comp_real-to-sim_2D}. Wedge cuts have been applied before averaging into 1D. In general, all instrumental effects add some systematic power, but the application of calibration exacerbates this. Source subtraction removes overall power in foreground-dominated modes, but does not help to correct instrumental errors.}
  \label{fig:comp_real-to-sim_1D_wedge}
\end{figure*}

The manifestations of each instrumental effect in isolation are shown in 1D PS in Figure~\ref{fig:comp_real-to-sim_1D_wedge}. In general, all instrumental effects add some systematic power, but the application of calibration exacerbates this. Source subtraction removes overall power in foreground-dominated modes, but does not help to correct instrumental errors. { In most cases, random errors add overall power, whereas systematic errors (e.g., flags, cable reflections) impart structured power.}

\subsection{Incomplete sky model}
\label{subsec:incomplete_skymodel}
In section \ref{subsec:pristine_direcsub}, completeness of the sky model used for subtraction revealed that discrete sources above an apparent flux density of 10$^{-3}$Jy needed to be removed to detect the \twocm signal. In addition, section \ref{subsec:pristine_calsub}, which used a set calibration sky model, showed that calibration alone can introduce significant power. Here we test whether, in the presence of instrumental effects, does increasing the number of calibrator sources improve leakage\? In this test, all dipoles are assumed to be functional (\hyperdrive only has to calculate one primary beam, and so it can do 15,000 calibration sources in about 15 minutes. With real data, we flag dead dipoles, and have to calculate numerous primary beam patterns, which is slower. So in these tests on real data, I have assumed all dipoles are alive. This might introduce a different calibration systematic.)

\begin{figure*}[h!]
  \centering
  \includegraphics[width=\columnwidth]{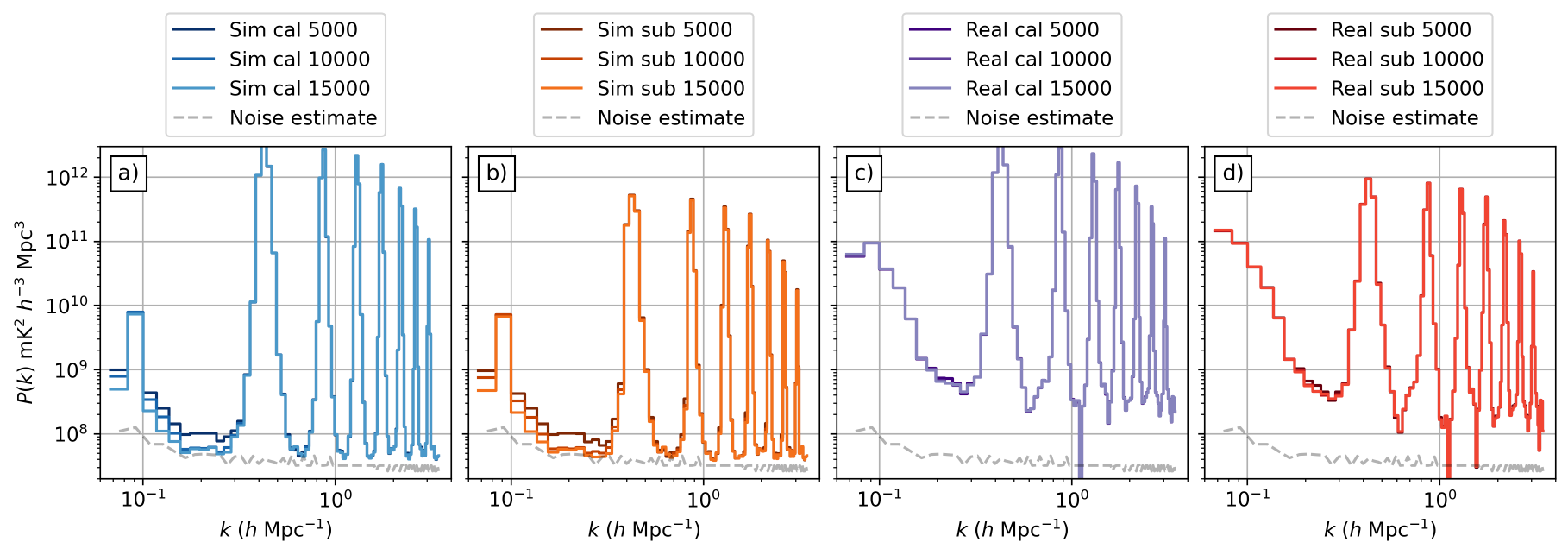}
  \caption{Comparison of calibrating and subtracting with various numbers of sources, with simulated data on left, real data on the right, when instrumental errors are included. Power spectra are made from integrating 15 EoR0 zenith observations. The simulations are noiseless as the effects of changing the calibration are below the noise threshold. Calibration was run on simulations including noise, so noise effects are carried into calibration solutions, and then applied to a noiseless simulation.}
  \label{fig:more_source_cal}
\end{figure*}

A full discrete + diffuse sky model is simulated, with 5,000, 10,000, or 15,000 sources used for calibration and subtraction. A matched set of 15 zenith EoR0 observations are then calibrated/subtracted with the same parameter sets. Figure \ref{fig:more_source_cal} displays the results as 1D power spectra. The simulations are noiseless as the effects of changing the calibration are below the noise threshold. Calibration was run on simulations including noise, so noise effects are carried into calibration solutions, and then applied to a noiseless simulation.
As observed previously, the simulated data outperform the real data, with significantly more systematic power in the latter. In the simulated datasets, inclusion of more calibration (and subtraction) sources, does reduce the leakage into the EoR window. Similar results are observed for the real data, but the relative amplitude of improvement is reduced compared with the simulations.

\section{Calibration averaging}
\label{sec:calib_averaging}
The results above included data simulated with thermal noise. In a 2-min snapshot, with calibration undertaken on 2-second cadence and 40$\,$kHz spectral resolution, the visibilities contain $\sim$30$\,$Jy of noise per channel, limiting the calibration precision. Ideally, one would want to average calibration solutions over time to reduce the noise, but this can only be achieved if the solutions are stable over time. The MWA has been shown to be a stable system (Jordan et al, 2024, \textit{submitted}), but changes in the beamformer settings to re-point the telescope can interrupt this. These individual pointings contain 15 observations (30 minutes long). It is reasonable to assume that solutions may be averaged over an individual pointing.

\begin{figure}[h!]
  \centering
  \includegraphics[width=\columnwidth]{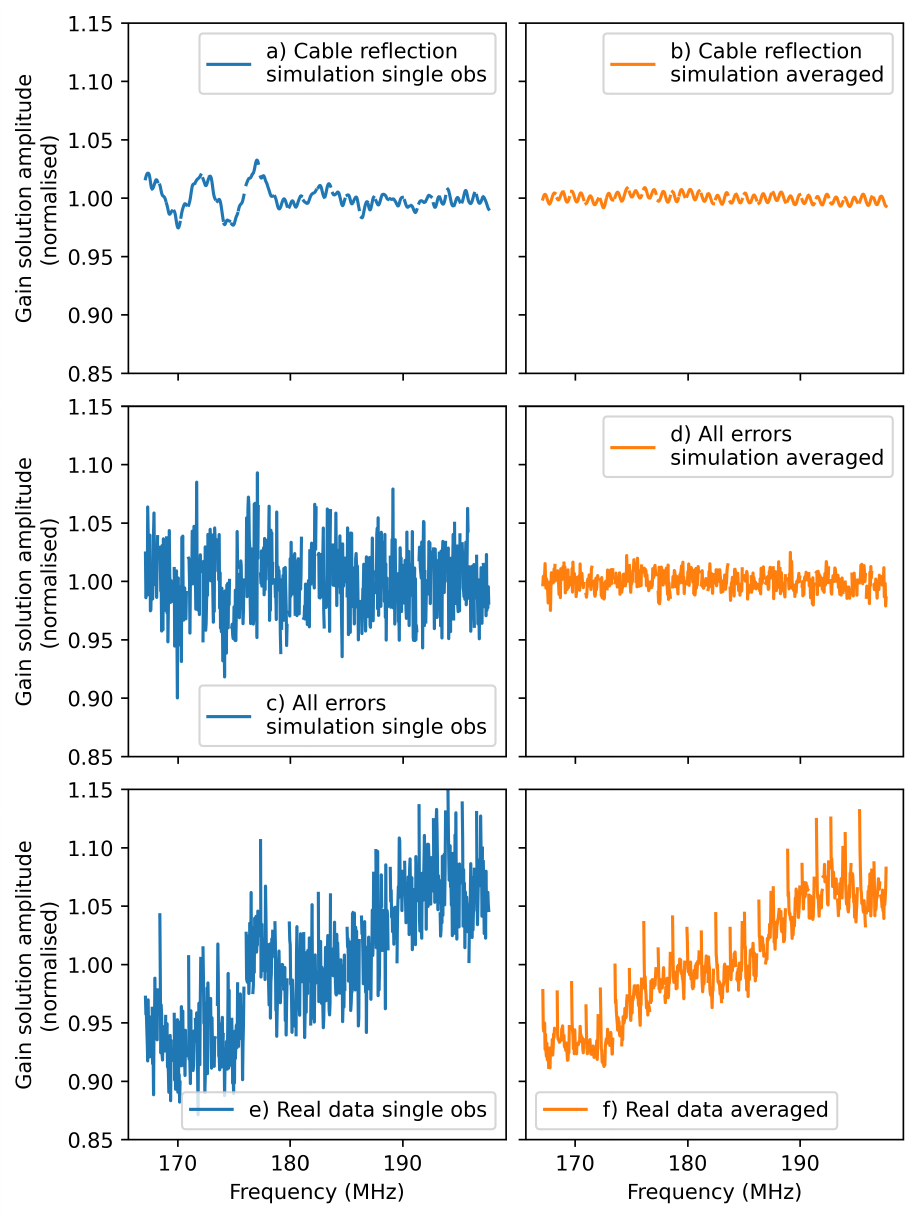}
  \caption{Calibration amplitudes showing the effects of averaging calibration solutions over 15 observations. Calibrations for a single tile are shown; other tiles show similar behaviours. Left column shows a single observation, right the average over 15 observations. Top row shows the simulation only containing cable reflection errors; middle row shows the simulation containing all instrumental errors; bottom shows real data.}
  \label{fig:avg_to_one_tile}
\end{figure}

Figure~\ref{fig:avg_to_one_tile} shows the effects of averaging for the calibration solution amplitudes of a single tile. The top row shows the only real frequency-dependent amplitude effect, which are cable reflections. The calibration errors coming from an incomplete sky model are obvious in Figure~\ref{fig:avg_to_one_tile}a, which seem to average out in Figure~\ref{fig:avg_to_one_tile}b to leave cable reflection ripples. The noisy calibration solutions obviously get less noisy with averaging.

\begin{figure*}[h!]
  \centering
  \includegraphics[width=0.94\columnwidth]{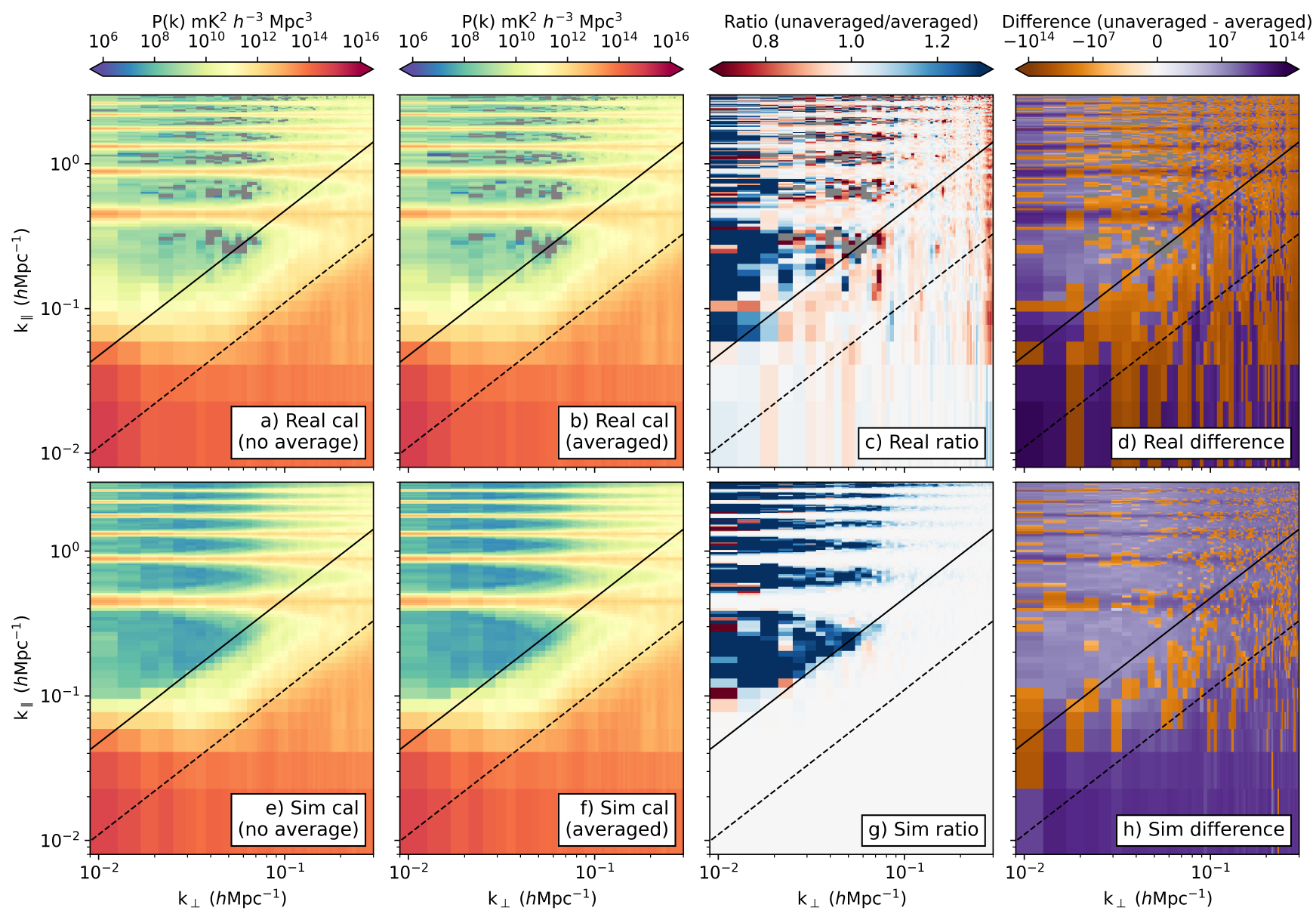}
  \caption{Effects of averaging calibration solutions over 15 observations for real data (top row) and simulated data (bottom row). The data in the simulation contain both discrete and diffuse sky models, but do not contain noise, in an attempt to better reveal any systematic bias involved in averaging. The calibration solutions applied to the simulation were derived from a simulation that did contain noise however, so the effects of noise are captured in the calibration solutions.}
  \label{fig:comp_avg_to_normal_cal_2D}
\end{figure*}

Although it is reasonable to average over a pointing, in reality there are complications like RFI, satellite passes, equipment failure etc that mean we may be averaging data with underlying signal differences, which could lead to bias. Figure~\ref{fig:comp_avg_to_normal_cal_2D} shows what happens when you average. Blue in the ratio (and purple in the difference means) less power in the window when you average the solutions, a.k.a less leakage. Of interest however is at low $k_\parallel, k_\perp$ in the real data (where the diffuse power resides), the averaged solutions result in less power. This can be seen by the blue in the ratio and purple in the difference in the bottom left pixels in  Figure~\ref{fig:comp_avg_to_normal_cal_2D}. It is hard to disentangle what is causing this reduction in power at large scales in the real data. Both simulations and real data show improvement in the EoR window with the averaging, which suggests that this is an avenue worth pursuing. Care must be taken in understanding what exactly is causing the reduction in both leakage and large-scale power for the real data. 

\section{Discussion}
\label{sec:discuss}
The work presented here is focussed on the MWA Australian pipeline, but many of the lessons are generally applicable to interferometers attempting experiments requiring high dynamic range. The interplay of different effects can exacerbate the impact of systematic errors, while some approaches are more well-suited to removing some systematics compared to others. For example, channel-based calibration handles cable reflections well, but is prone to imprinting spectral structure from calibration solutions on data, if smoothing is not used. A hybrid approach will be required. { A per-channel calibration strategy has the benefits of allowing large degrees of freedom for spectral structure, but at the cost of higher thermal noise. With an expectation that the pure instrumental gain response should have minimal spectral structure, one may take an intermediate approach where some spectral regularization of parametric fitting is undertaken. \citet{Li2019} used redundant calibration, along with a tile-based model for the structure imparted by cable reflections, whereas \citet{Yatawatta2009} and \citet{Yatawatta2010} used regularization to obtain smooth gain solutions across frequency. Some of these avenues have been explored with Hyperdrive, but none are currently used in production mode.}

\WODEN has been carefully designed to produce realistic MWA datasets, with many of the real instrumental effects observed in our data. The inclusion of diffuse emission in the simulations, currently an unused sky component in the MWA \hyperdrive calibration and subtraction pipeline, has demonstrated the significance of its effect on our ability to detect the \twocm signal. Despite the careful treatment and modelling of the full signal chain, there is clear evidence that there are unmodelled systematics that remain in the data (see, for example, Figure \ref{fig:comp_real-to-sim_2D}). The key conclusions that can be drawn from this work are:
\begin{itemize}
  \item With a sky-based calibration, uv-gridding and PS approach we have need to subtract more than 90\% of all discrete flux to recover \twocm signal in absence of instrumental effects;
  \item When including diffuse emission in simulations, we are never able to access some $k$-modes (from 30 minutes of data), leading to a need for some diffuse emission removal;
  \item The single greatest cause of leakage is an incomplete sky model;
  \item \hyperdrive easily treats simple tile gain errors and cable reflections;
\end{itemize}
The third point is worth discussion; without a higher-resolution and more sensitive southern hemisphere sky model, the MWA \textit{sky-based} calibration is currently limited to the mJy-level source detections available from LoBES and GLEAM-X. Early SKA arrays will have good sensitivity, but only array releases close to the full array will have the angular resolution and sensitivity increase needed to improve our discrete source sky model.
  
Based on the analysis presented here, there are likely three major updates needed in the \pipe for a detection:
  \begin{itemize}
      \item A gain smoothness enforcement in the calibration algorithm, similar to the approach of LOFAR \citep{Yatawatta2009}. This should intrinsically cause less leakage, but needs to be applied in a way that does not cause bias or signal loss;
      \item Fitting of cable reflections, similar to the FHD approach \citep{Beardsley2016}. This might allow averaging and/or fitting of calibration solutions to further reduce leakage;
      \item Treatment of the diffuse foregrounds. A foreground fitting approach, e.g. like GPR \citep{Mertens2020}, and in a way that is robust against \twocm signal loss.
  \end{itemize}

There are many options for future work to extend this analysis, including:
\begin{itemize}
  \item Increasing the field pointings that are simulated to match observations (i.e., away from zenith-pointed beams)
  \item Include frequency-dependent tile amplitude gain effects for more realistic instrumental effects
  \item Varying the short baseline cutoff to optimise for diffuse calibration
  \item Including full polarisation simulations and calibration
  \item Including the refractive effects of the ionosphere
  \item Including missing dipoles (different primary beams per tile) - this analysis is currently being undertaken
\end{itemize}

\section{Conclusion}
\label{sec:conclusion}
We used \WODEN, a full-sky simulator designed to produce realistic simulations of MWA EoR data, using discrete, \twocm, and diffuse sky sources. We added known direction-independent MWA instrumental effects to this simulated data. The simulations were used to test the impact of different effects on MWA data, and in particular, on the power spectrum of brightness temperature fluctuations of primordial hydrogen. We find that use of an incomplete sky model for data calibration has the largest impact on the ability to detect the EoR signal. Other effects such as channel and timestamp flagging, and cable reflections have less of an impact and do not prevent \twocm science. We also find that sources down to less than 10~mJy need to be removed for success, and that failing to treat diffuse emission removes signal accessibility for some modes.



\begin{acknowledgement}
This scientific work uses data obtained from \textit{Inyarrimanha Ilgari Bundara}, CSIRO's Murchison Radio-astronomy Observatory. We acknowledge the Wajarri Yamaji People as the Traditional Owners and native title holders of the Observatory site. Establishment of Inyarrimanha Ilgari Bundara is an initiative of the Australian Government, with support from the Government of Western Australia and the Science and Industry Endowment Fund. Support for the operation of the MWA is provided by the Australian Government (NCRIS), under a contract to Curtin University administered by Astronomy Australia Limited. This work was supported by resources provided by the Pawsey Supercomputing Research Centre with funding from the Australian Government and the Government of Western Australia. 

During this work we made extension use of the \texttt{kvis}~\citep{Gooch1996} and \texttt{DS9}~\citep{Joye2003} FITS file image viewers.
\end{acknowledgement}

\paragraph{Funding Statement}

This research was supported by the Australian Research Council Centre of Excellence for All Sky Astrophysics in 3 Dimensions (ASTRO 3D), through project number CE170100013. CMT is supported by an ARC Future Fellowship under grant FT180100321.

\paragraph{Data Availability Statement}
Simulated diffuse and discrete sky models can be made available upon request.

\printendnotes

\printbibliography

\appendix

\section{Calibration solutions}

\begin{figure*}[h!]
  \centering
  \includegraphics[width=\textwidth]{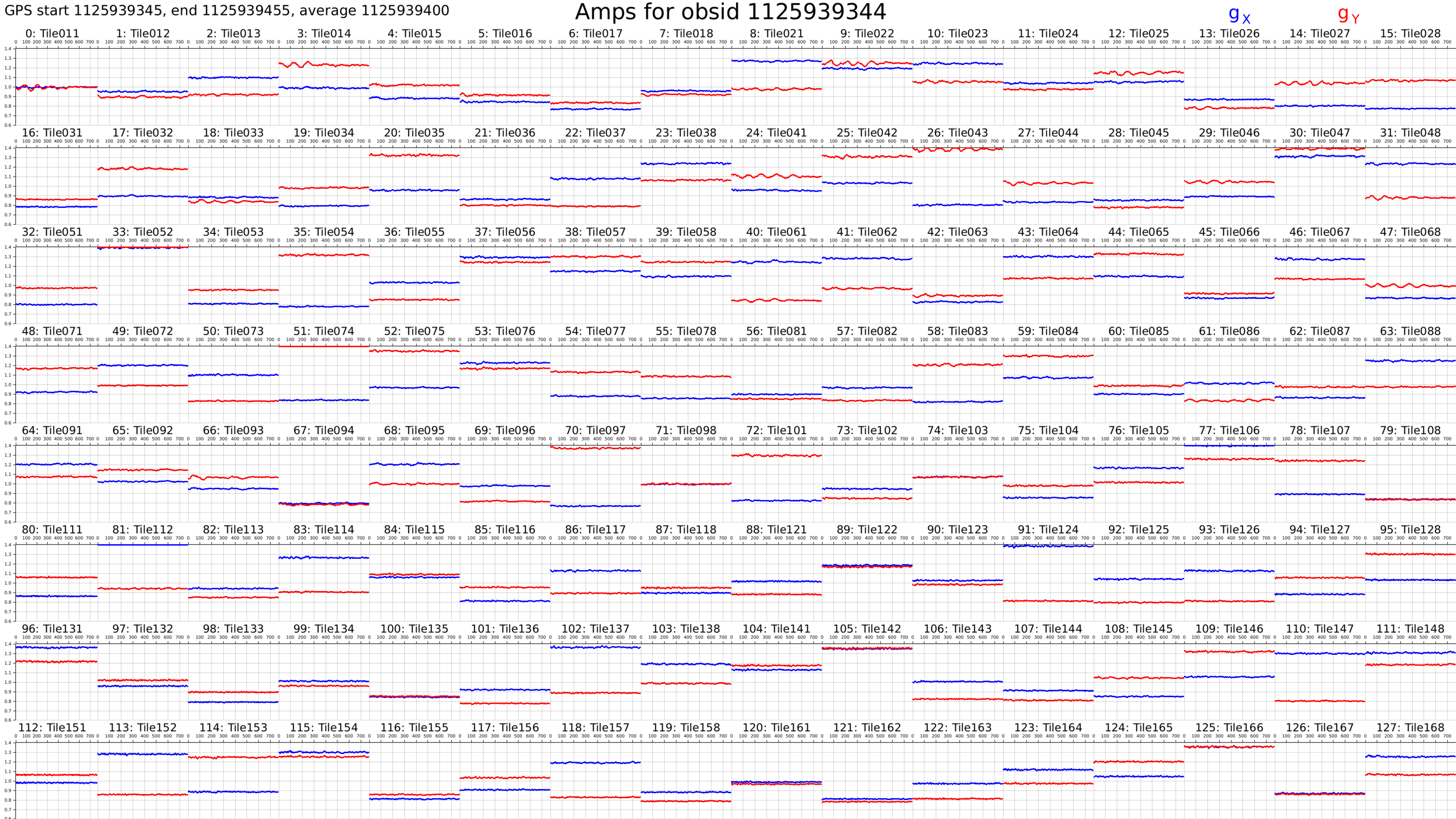}
  \includegraphics[width=\textwidth]{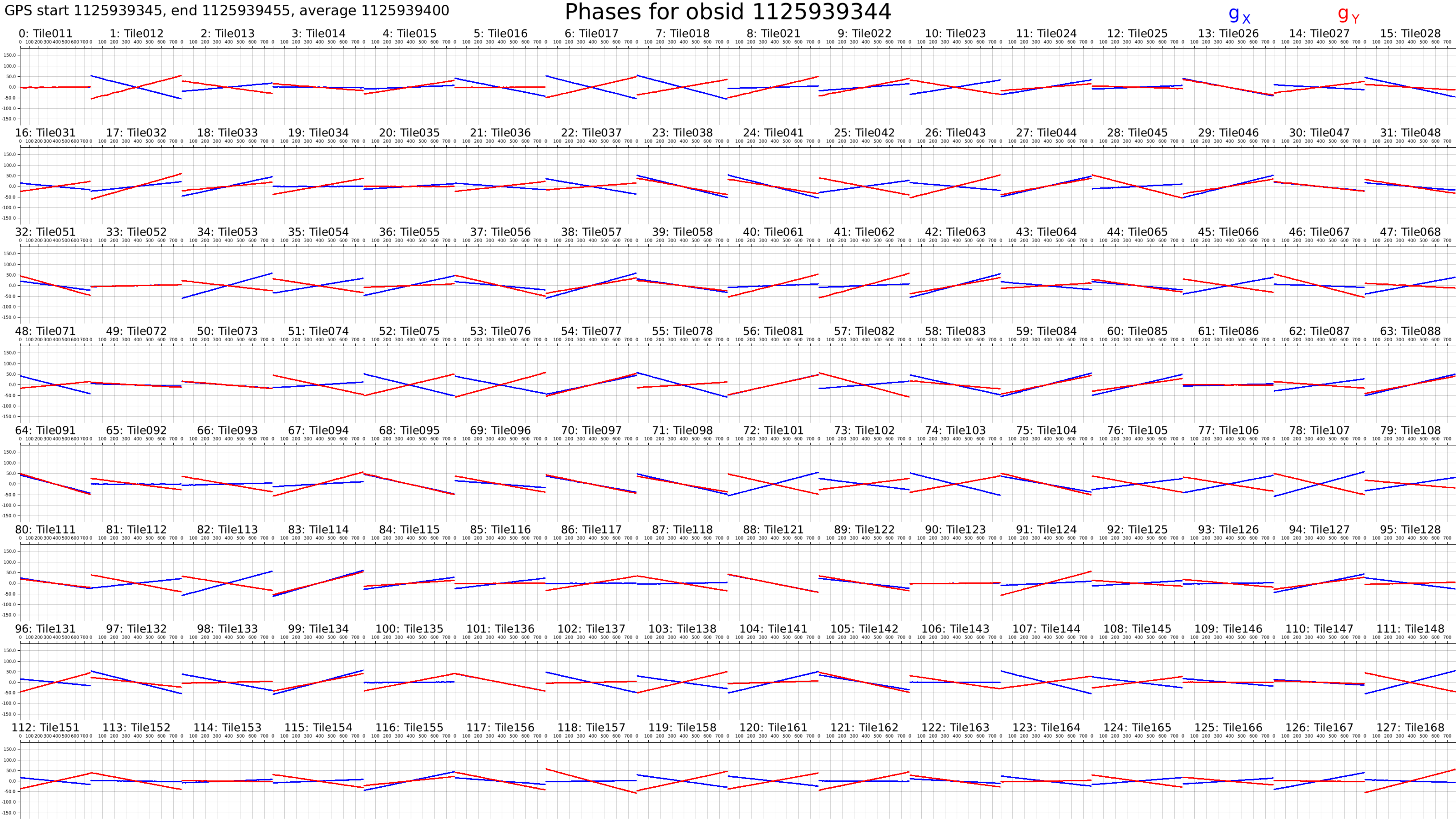}
  \caption{Calibration gain amplitudes and phases from a simulated two minute snapshot. These demonstrate the constant gain and flat phase slopes added to the simulation. The underlying simulation was of both the diffuse and discrete sky models, and only contained gain errors with no other instrumental effects. Calibration was performed using 10,000 sources through \hyperdrive. Any spectral structure in the amplitudes comes from the incomplete sky model rather than injected gains.}
  \label{appfig:gain_calsols}
\end{figure*}

\end{document}